\documentclass[12pt,letterpaper]{article}
\usepackage{jheppub}
\pdfoutput=1

\usepackage{amsmath, amsthm, amsfonts, amssymb}
\usepackage{graphicx}

\newcommand\sech {\mathrm{sech}}
\newcommand\csch {\mathrm{csch}}
\newcommand\am {\mathrm{am}}
\newcommand\sn {\mathrm{sn}}
\newcommand\cn {\mathrm{cn}}
\newcommand\dn {\mathrm{dn}}

\title{Revisiting the $O(3)$ Non-linear Sigma Model and Its Pohlmeyer Reduction}
\author[a,b]{Georgios Pastras}
\affiliation[a]{Department of Physics, School of Applied Mathematics and Physical Sciences, National Technical University, Athens 15780, Greece}
\affiliation[b]{NCSR ``Demokritos'', Institute of Nuclear and Particle Physics,
Aghia Paraskevi 15310, Greece}
\emailAdd{pastras@mail.ntua.gr}
\emailAdd{pastras@inp.demokritos.gr}

\abstract{It is well known that sigma models in symmetric spaces accept equivalent descriptions in terms of integrable systems such as the sine-Gordon equation through Pohlmeyer reduction. In this paper, we study the mapping between known solutions of the Euclidean $O(3)$ non-linear sigma model, such as instantons, merons and elliptic solutions that interpolate between the latter and solutions of the Pohlmeyer reduced theory, namely the sinh-Gordon equation. It turns out that instantons do not have a counterpart, merons correspond to the ground state, while the class of elliptic solutions is characterized by a two to one correspondence between solutions in the two descriptions.}

\keywords{Merons, Pohlmeyer Reduction}
\arxivnumber{}

\dedicated{In memoriam and gratitude of Prof. Ioannis Bakas}

\begin{document}

\maketitle


\section{Introduction}
\label{sec:Introduction}

Meron solutions of two-dimensional non-linear sigma models \cite{Gross:1977wu} are interesting singular solutions that are characterized by topological charge equal to $1/2$ and logarithmically divergent action. Instanton solutions in this class of models can be understood as being composed by two merons. These have been studied especially in the context of the relation of two-dimensional sigma models to four-dimensional Yang-Mills theories \cite{Polyakov:1975rs}, (for a review on the subject see \cite{Novikov:1984ac} and the references therein), which are also characterized by such solutions which dominate the path integral at strong coupling providing a qualitative picture for the confined phase of these theories \cite{Callan:1977qs}.

Elliptic solutions that interpolate between instanton and meron solutions have also been discovered \cite{Abbott:1982vf}. These solutions can be understood as the intermediate stages of the dissociation of meron pairs, supporting the relation between meron pairs and instantons. 

There is an alternative approach to study non-linear sigma models in symmetric spaces, the so called Pohlmeyer reduction \cite{Pohlmeyer:1975nb,Zakharov:1973pp,Eichenherr:1979yw,Pohlmeyer:1979ch,Eichenherr:1979ci,Eichenherr:1979hz,Eichenherr:1981pa}. The critical element of this approach is a non-local coordinate transformation that manifestly satisfies the Virasoro constraints, thus leaving only the physical degrees of freedom. In the case of the $O\left( 3 \right)$ model, this method leads to the relation of the sigma model to the sine-Gordon equation \cite{Pohlmeyer:1975nb}, while for higher dimensional symmetric spaces one results in multi-component generalizations of the latter.

Whether the reduced theory can be obtained from a local action is a non-trivial question in this approach, since the degrees of freedom of the reduced theory are connected to the original ones in a non-local way. It turns out that the reduced theory can be considered as a gauged WZW model with an integrable potential in a specific gauge \cite{Bakas:1993xh,Bakas:1995bm,FernandezPousa:1996hi}, allowing a systematic method to find the Lagrangian of the reduced theory.

The reduced theory has an integrable structure, allowing the construction of interesting solitonic solutions. However, the construction of the corresponding solution in the original sigma model is non-trivial due to the non-trivial relation between the original and reduced degrees of freedom. Such constructions have been of particular interest in the context of AdS/CFT correspondence in which case the non-linear sigma model describes the dynamics of strings propagating in AdS spaces \cite{Grigoriev:2007bu,Jevicki:2007aa,Klose:2008rx,Jevicki:2008mm,Rashkov:2008rm}.

Since several aspects of the relation between the sigma model and the reduced integrable theory have not been understood yet, mapping the solutions of the reduced theories to instanton, meron and elliptic solutions or the original sigma model may provide interesting insights to the properties of Pohlmeyer reduction.

The structure of this paper is as following: In section \ref{sec:NLSM} we review basic facts of the $O\left( 3 \right)$ non-linear sigma model. In section \ref{sec:NLSM_Solutions} we review the instanton, meron and elliptic solutions of the model. In section \ref{sec:Pohlmeyer} we study the Pohlmeyer reduction of the Euclidean $O\left( 3 \right)$ model resulting in the sinh-Gordon equation. In section \ref{sec:Reduced_Solutions}, we study the mapping between the solutions of the sigma model in the original formulation, as reviewed in section \ref{sec:NLSM}, and solutions of the sinh-Gordon equation. Finally, in section \ref{sec:Discussion} we discuss our results. There is also en appendix with properties of Jacobi elliptic functions that are used throughout the text.

\section{The Euclidean $O(3)$ Non-linear Sigma Model}
\label{sec:NLSM}
\subsection{The Model}
\label{subsec:NLSM_model}
The Euclidean $O(3)$ non-linear sigma model describes the dynamics of a three component vector field $X^i \left(\sigma_0 , \sigma_1 \right)$ with unit norm. Its action is given by
\begin{equation}
\mathcal{S} = \frac{1}{{{g^2}}}\int {d{x_0}d{x_1}\left( {\frac{1}{2}{\partial _m}{X^i}{\partial _m}{X^i} - 2\lambda \left( {{X^i}{X^i} - 1} \right)} \right)} .
\label{eq:initial_action}
\end{equation}
Index $i$ takes the values 1, 2, 3, while index $m$ takes the values 0, 1 and is contracted with the Euclidean metric $\delta_{mn}$. We first introduce the complex coordinate $z$,
\begin{equation}
z = {x_0} + i{x_1} .
\end{equation}
Then, the action of the Euclidean $O(3)$ model is written as
\begin{equation}
\mathcal{S} = \frac{1}{{{g^2}}}\int {dzd\bar z\left( {\partial {X^i}\bar \partial {X^i} - \lambda \left( {{X^i}{X^i} - 1} \right)} \right)} .
\label{eq:action_X}
\end{equation}
The constraint can be easily satisfied introducing angular field variables $\Theta$ and $\Phi$ as
\begin{align}
{X^1} &= \sin \Theta \cos \Phi , \label{eq:angular_fields_definition_1} \\
{X^2} &= \sin \Theta \sin \Phi , \label{eq:angular_fields_definition_2} \\
{X^3} &= \cos \Theta . \label{eq:angular_fields_definition_3}
\end{align}
Then, the action is written as
\begin{equation}
\mathcal{S} = \frac{1}{{{g^2}}}\int {dzd\bar z\left( {\partial \Theta \bar \partial \Theta  + {{\sin }^2}\Theta \partial \Phi \bar \partial \Phi } \right)} .
\label{eq:action_angular}
\end{equation}
Finally, we can define the ``stereographic projection'' complex field variable $W$ as,
\begin{equation}
W = \cot \frac{\Theta }{2}{e^{i\Phi }} .
\label{eq:stereographic_field_definition}
\end{equation}
and the action is written as
\begin{equation}
\mathcal{S} = \frac{2}{{{g^2}}}\int {dzd\bar z\frac{{\partial W\bar \partial \bar W + \partial \bar W\bar \partial W}}{{{{\left( {1 + W\bar W} \right)}^2}}}} .\label{eq:action_stereographic}
\end{equation}

Any field configuration describes a mapping from the complex plane to the sphere $S^2$. Any finite action configuration must have a unique  limit as the magnitude of the complex number $z$ tends to infinity. In such cases, the complex plane can be compactified to the sphere $S^2$ and the finite action field configurations are mappings from $S^2$ to $S^2$, thus, they are classified by the homotopy classes of these mappings, which are trivially the elements of the group of integers $\mathbb{Z}$. One can define the topological number density as the Jacobian of the mapping
\begin{equation}
\mathcal{Q} = \frac{1}{{8\pi }}{\varepsilon ^{ijk}}{\varepsilon _{mn}}{X^i}{\partial _m}{X^j}{\partial _n}{X^k} = \frac{1}{\pi } \frac{{\partial W\bar \partial \bar W - \partial \bar W\bar \partial W}}{{{{\left( {1 + W\bar W} \right)}^2}}} .\label{eq:Topological_Charge_W}
\end{equation}

\subsection{The Equations of Motion and the Stress-energy Tensor}
\label{subsec:NLSM_eom}

It will turn out that all the formulations of the Euclidean $O(3)$ non-linear sigma model presented in section \ref{subsec:NLSM_model}, given by equations \eqref{eq:action_X}, \eqref{eq:action_angular} and \eqref{eq:action_stereographic} will be useful for different points of view of the model solutions. For this reason we will derive the equations of motion and the stress-energy tensor in all these formulations.

\subsubsection*{Scalar Fields Triplet Formulation}
Starting from action \eqref{eq:action_X}, the equations of motion for the fields $X^m$ are
\begin{equation}
{\partial }{\bar \partial }{X^i} + \lambda {X^i} = 0 ,
\label{eq:eom_X_1}
\end{equation}
while the equation of motion for the Lagrange multiplier $\lambda$ is the constraint equation
\begin{equation}
{X^i}{X^i} = 1 .
\end{equation}
Acting with ${\partial }{\bar \partial }$ on the constraint equation we get
\begin{equation}
{\partial }{X^i}{\bar \partial }{X^i} + {X^i}{\partial }{\bar \partial }{X^i} = 0 .
\end{equation}
The latter, using both equations of motion for $X^i$ and the constraint, can be written as
\begin{equation}
\lambda  = {\partial }{X^i}{\bar \partial }{X^i} ,
\end{equation}
implying that we can write down the equations of motion for the fields $X^i$ decoupled from the Lagrange multiplier as
\begin{equation}
{\partial }{\bar \partial }{X^i} + \left( {\partial }{X^j}{\bar \partial }{X^j} \right) {X^i} = 0 .
\label{eq:eom_X_2}
\end{equation}

We can also calculate the stress-energy tensor and find it equal to
\begin{align}
{T_{zz}} &= \frac{1}{g^2} \partial X^i \partial X^i ,\label{eq:Tzz_X}\\
{T_{\bar z\bar z}} &= \frac{1}{g^2} \bar \partial X^i \bar \partial X^i ,\label{eq:Tzbzb_X}\\
{T_{z \bar z}} &= 0 .\label{eq:Tzzb_X}
\end{align}

\subsubsection*{Angular Fields Formulation}

The equations of motion derived from action \eqref{eq:action_angular} are
\begin{align}
\partial \bar \partial \Theta  + \sin 2 \Theta \left( {\partial \Phi \bar \partial \Phi } \right) &= 0\\
\partial \bar \partial \Phi  &= 0
\end{align}

The stress-energy tensor in the angular field variables formulation equals
\begin{align}
{T_{zz}} &= \frac{1}{g^2} \left[{\left( {\partial \Theta } \right)^2} + {\sin ^2}\Theta {\left( {\partial \Phi } \right)^2} \right] , \label{eq:Tzz_angular} \\
{T_{\bar z\bar z}} &= \frac{1}{g^2} \left[ {\left( {\bar \partial \Theta } \right)^2} + {\sin ^2}\Theta {\left( {\bar \partial \Phi } \right)^2} \right] , \label{eq:Tzbzb_angular} \\
{T_{z\bar z}} &= 0 \label{eq:Tzzb_angular} .
\end{align}

\subsubsection*{Stereographic Complex Field Formulation}

The equations of motion derived from action \eqref{eq:action_stereographic} are
\begin{equation}
\left( {1 + W\bar W} \right)\partial \bar \partial W - 2\bar W\partial W\bar \partial W = 0 .\label{eq:eom_W}
\end{equation}

The stress-energy tensor equals
\begin{align}
{T_{zz}} &= \frac{4}{{{g^2}}} \frac{{\partial W\partial \bar W}}{{{{\left( {1 + W\bar W} \right)}^2}}} ,\label{eq:Tzz_W}\\
{T_{\bar z\bar z}} &= \frac{4}{{{g^2}}} \frac{{\bar \partial W\bar \partial \bar W}}{{{{\left( {1 + W\bar W} \right)}^2}}} ,\label{eq:Tzbzb_W}\\
{T_{z \bar z}} &= 0 .\label{eq:Tzzb_W}
\end{align}

As expected, ${T_{z \bar z}}$ vanishes identically as a result of conformal symmetry in all three formulations.

\section{Solutions of the $O(3)$ Non-linear Sigma Model}
\label{sec:NLSM_Solutions}

The $O(3)$ non-linear Sigma Model is known to have several interesting solutions. We will distinguish three classes of solutions: solutions of integer topological charge, i.e. instantons, solutions of half-integer topological charge, i.e. merons and solutions that interpolate between merons and instantons. For the analysis in this section, we will use the formulation of the sigma model in terms of the complex field variable $W$.

\subsection{Instantons}
\label{subsec:NLSM_instantons}

Clearly, any holomorphic function $W$ or antiholomorphic function $\tilde W$
\begin{align}
\bar \partial W &= 0 , \label{eq:instantons}\\
\partial \tilde W &= 0 \label{eq:antiinstantons}
\end{align}
is a solution of the equations of motion \eqref{eq:eom_W}. If one wants to restrict to solutions that are characterized by finite action, i.e. solutions that become single valued at infinity, the general solution has to be written as product of factors of the form $\frac{{z - {a^ - }}}{{z - {a^ + }}}$, for instantons and as product of factors of the form $\frac{{\bar z - {a^ - }}}{{\bar z - {a^ + }}}$, for antiinstantons \cite{Polyakov:1975yp},
\begin{align}
W &= \prod\limits_{i = 1}^N {\frac{{z - a_i^ - }}{{z - a_i^ + }}} , \label{eq:multiinstantons}\\
\tilde W &= \prod\limits_{i = 1}^N {\frac{{\bar z - a_i^ - }}{{\bar z - a_i^ + }}} . \label{eq:multiantiinstantons}
\end{align}

It is easy to show that upon substitution of \eqref{eq:multiinstantons} and \eqref{eq:multiantiinstantons} to the expression for the topological charge \eqref{eq:Topological_Charge_W}, we find
\begin{align}
Q^{\left( {\mathrm{inst}} \right)} &= N , \\
Q^{\left( {\mathrm{antiinst}} \right)} &= - N .
\end{align}
Studying the form of the action for the solutions \eqref{eq:multiinstantons} and \eqref{eq:multiantiinstantons} and having found the corresponding topological charges, it is natural to interpret these solutions as $N$-instanton or $N$-antiinstanton solutions respectively, localized at positions $\frac{1}{2} \left( a_i^ + + a_i^ - \right)$. Instantons and antiinstantons are the only finite action solutions of the sigma model \cite{Woo:1977gi}. Further details are beyond the scope of this work.

Equations \eqref{eq:Tzz_W} and \eqref{eq:Tzbzb_W} imply that the instanton and anti-instanton solutions are characterized by vanishing stress-energy tensor.
\begin{align}
T_{zz}^{\left( {\mathrm{inst}} \right)} = T_{\bar z\bar z}^{\left( {\mathrm{inst}} \right)} &= 0 , \\
T_{zz}^{\left( {\mathrm{antiinst}} \right)} = T_{\bar z\bar z}^{\left( {\mathrm{antiinst}} \right)} &= 0 .
\end{align}

\subsection{Merons}
\label{subsec:NLSM_merons}

Meron solutions are solutions characterized by vanishing topological charge except for singular points. Since the topological charge density is simply the Jacobian of the mapping defined by the fields $X^i$, merons correspond to degenerate solutions that map the complex plane to one-dimensional submanifolds of the sphere. Demanding the above, it turns out that the general meron solution of equations \eqref{eq:eom_W} is of the form \cite{Gross:1977wu}
\begin{equation}
W = \sqrt {\frac{{f\left( z \right)}}{{\bar f\left( {\bar z} \right)}}} , 
\label{eq:meron_solution}
\end{equation}
where $f\left( z \right)$ is an arbitrary meromorphic function of $z$. Furthermore, the demand that the solution is single valued at infinity results in a specific selection of function $f\left( z \right)$,
\begin{equation}
W = {\left( {\prod\limits_{i = 1}^N {\frac{{z - a_i^ - }}{{\bar z - \bar a_i^ - }}\frac{{\bar z - \bar a_i^ + }}{{z - a_i^ + }}} } \right)^{\frac{1}{2}}} ,
\label{eq:multimerons}
\end{equation}
which corresponds to a $2N$ meron configuration, with merons localized at $ a_i^ -$ and $ a_i^ +$. For more details on merons, the reader is encouraged to read \cite{Gross:1977wu}.

Substituting \eqref{eq:meron_solution} to \eqref{eq:Tzz_W} and \eqref{eq:Tzbzb_W}, it is simple to find that the stress-energy tensor elements corresponding to the solution \eqref{eq:meron_solution} are
\begin{align}
{T_{zz}} &= - \frac{1}{4 g^2} {\left( {\frac{{f'\left( z \right)}}{{f\left( z \right)}}} \right)^2} ,\label{eq:meron_solution_Tzz}\\
{T_{\bar z\bar z}} &= - \frac{1}{4 g^2} {\left( {\frac{{\bar f'\left( {\bar z} \right)}}{{\bar f\left( {\bar z} \right)}}} \right)^2} .\label{eq:meron_solution_Tzbzb}
\end{align}

\subsection{Elliptic Solutions}
\label{subsec:Elliptic}

Although, the instanton and meron solutions are both infinite families of solutions, many more solutions have been neglected due to the demand of localized topological charge. In this section, we will abandon this demand, and following \cite{Abbott:1982vf}, we will find solutions interpolating between the single meron solution $W = \sqrt {z /{\bar z}}$ and the single instanton solution $W = z$. To do so, we substitute in equations \eqref{eq:eom_W} the ansatz
\begin{equation}
W = \sqrt {\frac{z}{{\bar z}}} \tan \frac{{\psi \left( t \right)}}{4} ,
\label{eq:ansatz}
\end{equation}
where
\begin{equation}
t = \ln \left| z \right| .
\end{equation}
It is a matter of simple algebra to find that
substituting \eqref{eq:ansatz} in equations \eqref{eq:action_stereographic}, \eqref{eq:Topological_Charge_W}, \eqref{eq:Tzz_W} and \eqref{eq:Tzbzb_W}, yields
\begin{align}
S &= \frac{\pi }{{{2 g^2}}}\int {dt\left( {\frac{1}{2}{{\left( {\frac{{d\psi }}{{dt}}} \right)}^2} + \left( {1 - \cos \psi } \right)} \right)} ,\label{eq:ansatz_L} \\
Q &= \frac{1}{2}\int {dt\frac{d}{{dt}}\left( {\cos \frac{\psi }{2}} \right)} \label{eq:ansatz_Q}
\end{align}
and
\begin{align}
{T_{zz}} &= \frac{1}{{{8 g^2 z^2}}}\left( {\frac{1}{2}{{\left( {\frac{{d\psi }}{{dt}}} \right)}^2} - \left( {1 - \cos \psi } \right)} \right) ,\label{eq:ansatz_Tzz}\\
{T_{\bar z\bar z}} &= \frac{1}{{8 g^2 {{\bar z}^2}}}\left( {\frac{1}{2}{{\left( {\frac{{d\psi }}{{dt}}} \right)}^2} - \left( {1 - \cos \psi } \right)} \right) .\label{eq:ansatz_Tzbzb}
\end{align}

Equation \eqref{eq:ansatz_L} implies that the unspecified function $\psi \left( t \right)$ obeys the equation
\begin{equation}
\frac{{{d^2}\psi }}{{d{t^2}}} = \sin \psi ,\label{eq:ansatz_eom}
\end{equation}
which is the one-dimensional sine-Gordon equation or in other words the equation of motion of the pendulum with time having been substituted with the spacial variable $t$ and potential equal to
\begin{equation}
V = \cos \psi  - 1 .\label{eq:ansatz_potential}
\end{equation}
Thus, this effective one-dimensional problem has a stable equilibrium that lies at $\psi = \pi$ and an unstable one at $\psi = 0$.

It has to be pointed out that this one-dimensional sine-Gordon has nothing to do with the sine-Gordon that occurs after the Pohlmeyer reduction\footnote{Actually, this is the case for the O$(3)$ NLSM with Minkowski signature. In section \ref{sec:Pohlmeyer}, we will show that the Euclidean O$(3)$ NLSM is mapped through Pohlmeyer reduction to the sinh-Gordon equation.}. Unlike the connection between the NLSM degrees of freedom and the Pohlmeyer reduced field, which is non-local, as we will see in section \ref{sec:Pohlmeyer}, in this special case $\psi$ is locally connected to the degrees of freedom of the NLSM through \eqref{eq:ansatz}. This kind of mapping of the O$(3)$ NLSM to the sine-Gordon equation is a property of the particular model only and it greatly facilitates the study of its elliptic solutions.

Simple solutions of this effective problem can give us an idea about how the general solution of this problem interpolates between the single meron and single instanton solutions. The unstable vacuum solution $\psi = 0$ corresponds to the trivial solution $W = 0$. The kink solution $\psi  = 4{\tan ^{ - 1}}{e^t} = 4{\tan ^{ - 1}}\left| z \right|$ corresponds to the one-instanton solution $W = z$. Finally the stable vacuum solution $\psi = \pi$ corresponds to the one-meron solution $W = \sqrt {z / \bar z}$. Thus, the general solution to be found should have the limits of the stable vacuum solution and the kink solution.

It is a direct consequence of the equation of motion \eqref{eq:ansatz_eom} that
\begin{equation}
\frac{1}{2}{\left( {\frac{{d\psi }}{{dt}}} \right)^2} + \left( {\cos \psi  - 1} \right) = E .\label{eq:ansatz_energy_conservation}
\end{equation}
In the language of the pendulum the above would be simply energy conservation. We can categorize the solutions in the following classes
\begin{itemize}
\item $E = - 2$ Stable equilibrium
\item $- 2 < E < 0$ Periodic solutions
\item $E = 0$ Unstable equilibrium and kink solutions
\item $E > 0$ Quasi-periodic solutions
\end{itemize}

The integral \eqref{eq:ansatz_energy_conservation} of the equations of motion, allows the expression of the general solution in terms of the Jacobi elliptic functions (see also \cite{Bakas:2002qi}). Several useful definitions and properties of Jacobi elliptic functions are summarized in appendix \ref{sec:Elliptic_formulas}. If one defines $\psi  = 2\Psi  + \pi $, equation \eqref{eq:ansatz_energy_conservation} is written as
\begin{equation}
{\left( {1 - \frac{2}{{E + 2}}{{\sin }^2}\Psi } \right)^{ - \frac{1}{2}}}d\Psi  = {\left( {\frac{{E + 2}}{2}} \right)^{\frac{1}{2}}}dt .
\end{equation}
Assuming an initial condition of the form $\psi \left( t_0 \right) = \pi$ (or $\Psi \left( t_0 \right) = 0$), the above expression can be integrated using the definition of the incomplete elliptic integral of the first kind \eqref{eq:Elliptic_integral}
\begin{equation}
F\left( {\Psi ;\left( \frac{2}{{E + 2}} \right)^{\frac{1}{2}} } \right) = {\left( {\frac{{E + 2}}{2}} \right)^{\frac{1}{2}}}\left( {t - {t_0}} \right) ,
\end{equation}
or
\begin{equation}
\Psi  = \am\left( {{{\left( {\frac{{E + 2}}{2}} \right)}^{\frac{1}{2}}}\left( {t - {t_0}} \right);\left( \frac{2}{{E + 2}} \right)^{\frac{1}{2}}} \right) ,
\end{equation}
where $F\left( x;k \right)$ is the incomplete elliptic integral of the first kind with modulus $k$ and $\am\left( x;k \right)$ is its inverse function, namely the amplitude of Jacobi elliptic functions. It is a matter of simple algebra to find that
\begin{equation}
\begin{split}
W^{\left( qp \right)}\left(z,\bar z;E \right) &= \sqrt {\frac{z}{{\bar z}}} \tan \left[ {\frac{\pi }{4} + \frac{1}{2}\am\left( {{{\left( {\frac{{E + 2}}{2}} \right)}^{\frac{1}{2}}}\left( {t - {t_0}} \right);\left( \frac{2}{{E + 2}} \right)^{\frac{1}{2}}} \right)} \right]\\
 &= \sqrt {\frac{z}{{\bar z}}} \sqrt {\frac{{1 + \sn\left( {{{\left( {\frac{{E + 2}}{2}} \right)}^{\frac{1}{2}}}\left( {t - {t_0}} \right);\left( \frac{2}{{E + 2}} \right)^{\frac{1}{2}}} \right)}}{{1 - \sn\left( {{{\left( {\frac{{E + 2}}{2}} \right)}^{\frac{1}{2}}}\left( {t - {t_0}} \right);\left( \frac{2}{{E + 2}} \right)^{\frac{1}{2}}} \right)}}} .
\end{split}
\label{eq:elliptic_solution_quasi-periodic}
\end{equation}
For $E<0$ the modulus of the Jacobi elliptic functions does not lie in the interval $\left[0,1\right]$. However, use of appropriate properties of Jacobi elliptic functions \eqref{eq:sn_reciprocal} allows relating these quantities to elliptic functions with modulus in the usual interval as
\begin{equation}
W^{\left( p \right)}\left(z,\bar z;E \right) = \sqrt {\frac{z}{{\bar z}}} \sqrt {\frac{{1 + {{\left( {\frac{{E + 2}}{2}} \right)}^{\frac{1}{2}}}\sn\left( {\left( {t - {t_0}} \right);{{\left( {\frac{{E + 2}}{2}} \right)}^{\frac{1}{2}}}} \right)}}{{1 - {{\left( {\frac{{E + 2}}{2}} \right)}^{\frac{1}{2}}}\sn\left( {\left( {t - {t_0}} \right);{{\left( {\frac{{E + 2}}{2}} \right)}^{\frac{1}{2}}}} \right)}}} .
\label{eq:elliptic_solution_periodic}
\end{equation}

As physically expected from the pendulum mechanical analog of the problem, the above solutions are periodic for $E < 0$, and quasi-periodic for $E > 0$, thus the indices in expressions \eqref{eq:elliptic_solution_quasi-periodic} and \eqref{eq:elliptic_solution_periodic}. Periodic properties of Jacobi elliptic functions \eqref{eq:sn_period} allow the specification of the corresponding periods,
\begin{equation}
\begin{cases}
T_p = 4 K \left( \sqrt{\frac{E + 2}{2}} \right) , & E < 0 ,\\
T_{qp} = 2 \sqrt{\frac{2}{E + 2}} K \left( \sqrt{\frac{2}{E + 2}} \right) , & E > 0 ,
\end{cases}
\label{eq:elliptic_solution_period}
\end{equation}
where $K\left( x \right)$ is the complete elliptic integral of the first kind.

Property \eqref{eq:zero_argument_sn} of Jacobi elliptic functions implies that at the limit $E \to -2$ one recovers the single meron solution
\begin{equation}
\mathop {\lim }\limits_{E \to  - 2} {W^{\left( p \right)}}\left( {z,\bar z;E} \right) = \sqrt {\frac{z}{{\bar z}}}  = {W^{\left( {\mathrm{meron}} \right)}} ,
\end{equation}
as expected. On the other hand, property \eqref{eq:unit_modulus_sn} of Jacobi elliptic functions implies that at the limit $E \to 0$ both periodic and quasi-periodic solutions tend to the instanton solution
\begin{equation}
\mathop {\lim }\limits_{E \to 0} {W^{\left( p \right)}}\left( {z,\bar z;E} \right) = \mathop {\lim }\limits_{E \to 0} {W^{\left( {qp} \right)}}\left( {z,\bar z;E} \right) = z = {W^{\left( {\mathrm{inst}} \right)}} .
\end{equation}

\begin{figure}[h]
\begin{center}
\includegraphics[width=0.4\textwidth]{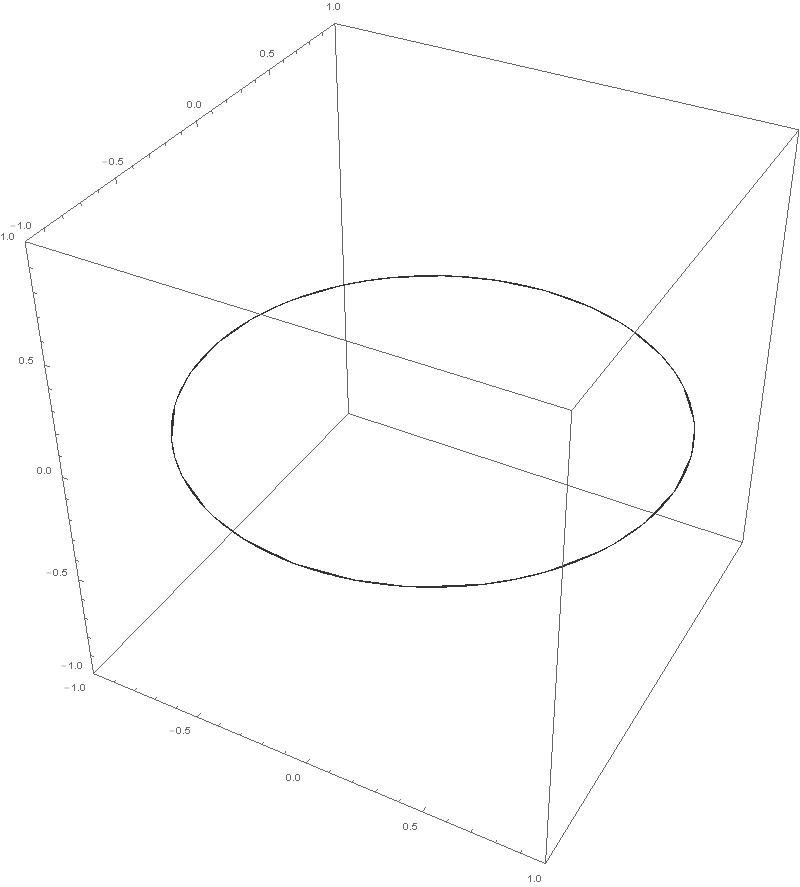}\includegraphics[width=0.4\textwidth]{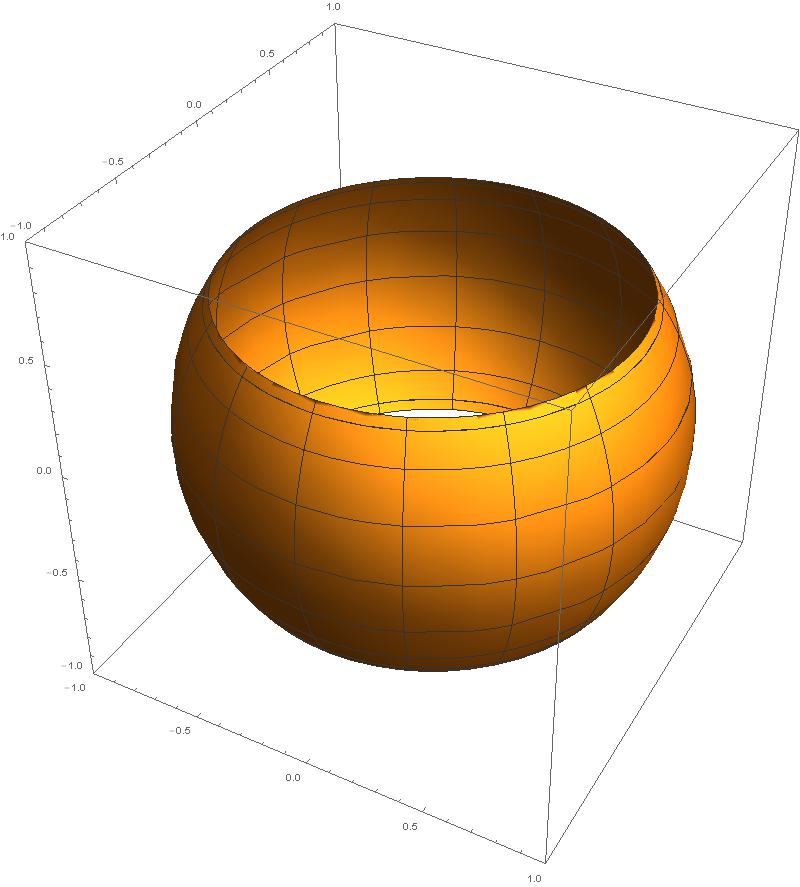}\\
\includegraphics[width=0.4\textwidth]{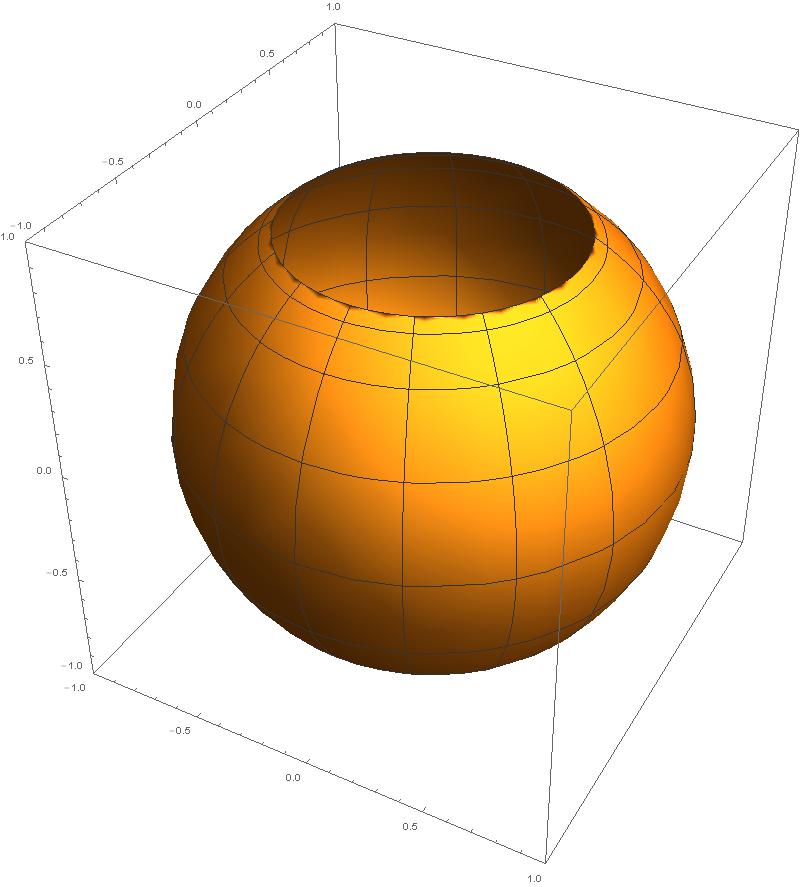}\includegraphics[width=0.4\textwidth]{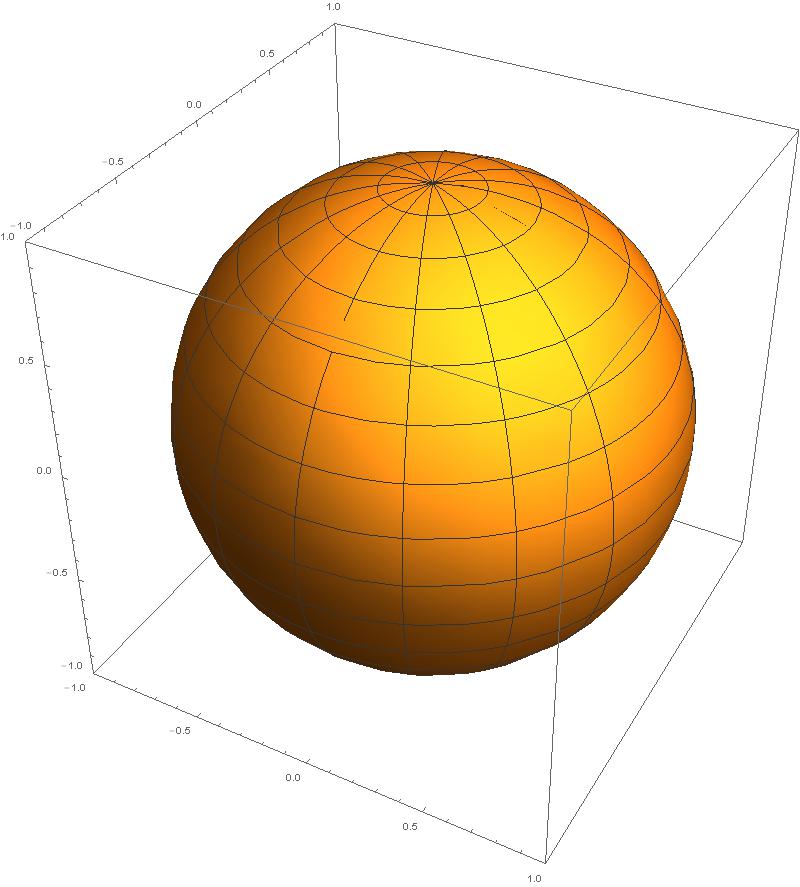}
\end{center}
\caption{The surfaces in $S^2$ corresponding to the elliptic solutions and the meron and instanton limits. Top-left: $E=-2$ (meron solution), top-right: $E=-4/3$, bottom-left: $E=-2/3$, bottom-right: $E=0$ (instanton solution)}
\label{fig:elliptic_solutions_surfaces}
\end{figure}
In figure \ref{fig:elliptic_solutions_surfaces}, the surfaces in $S^2$ corresponding to the periodic elliptic solutions and the meron and instanton limits are displayed. It can be seen that indeed the elliptic solutions interpolate between the degenerate mapping of the meron configuration to the identity mapping from $S^2$ to $S^2$ that corresponds to a single instanton.

We would also like to comment that the topological charge, as implied by equation \eqref{eq:ansatz_Q}, is not well defined for the elliptic solutions due to the non-existence of the limit $\mathop {\lim }\limits_{t \to \infty } \psi \left( t \right)$. Of course at the limits $E \to -2$ and $E \to 0$, corresponding to the stable equilibrium and kink solutions respectively, the solution is neither periodic nor quasi-periodic and the limit $\mathop {\lim }\limits_{t \to \infty } \psi \left( t \right)$ is well defined, giving rise to a well defined topological charge number, as required for meron and instanton solutions.

For solutions \eqref{eq:elliptic_solution_quasi-periodic} and \eqref{eq:elliptic_solution_periodic}, as can be easily seen from equations \eqref{eq:ansatz_Tzz} \eqref{eq:ansatz_Tzbzb}, the existence of the integral \eqref{eq:ansatz_energy_conservation} implies that the stress-energy tensor components are equal to
\begin{align}
T_{zz} &= \frac{E}{8 g^2 z^2} , \label{eq:elliptic_solution_Tzz}\\
T_{\bar z \bar z} &= \frac{E}{8 g^2 {\bar z}^2} . \label{eq:elliptic_solution_Tzbzb}
\end{align}
It is easy to check that the above expressions agree with the fact that the instanton solutions are characterized by vanishing stress-energy tensor, as well the formulas \eqref{eq:meron_solution_Tzz} and \eqref{eq:meron_solution_Tzbzb}, giving the stress-energy tensor components for a meron configuration at the appropriate limit $E \to -2$ and $f\left( z \right) \to z$.

\pagebreak
\section{Reduction of the Euclidean $O(3)$ Non-linear Sigma Model}
\label{sec:Pohlmeyer}

The Pohlmeyer reduced theory is derived based on the formulation of the sigma model in terms of the scalar fields triplet $X^i$, that take values in a symmetric target space, namely $S^2$. The most known example of such reduction is for the Lorentzian version of the model under study, leading to the sine-Gordon equation \cite{Pohlmeyer:1975nb}. Here we will follow closely the analysis of \cite{Pohlmeyer:1975nb} in the reduction of the Euclidean version of $O(3)$ non-linear sigma model.

\subsection{Definition of the Reduced Theory Field}
\label{subsec:Pohlmeyer_definitions}

First, we would like to observe that by definition of the $O(3)$ sigma model, the scalar fields obey the constraint ${X^i}{X^i} = 1$. It is a direct consequence that the vector $X^i$ is perpendicular to the vectors $\partial X^i$ and $\bar \partial X^i$.
\begin{equation}
\left( {\partial }{X^i} \right) {X^i} = \left( {\bar \partial }{X^i} \right) {X^i} = 0.
\label{eq:X_orthogonal_to_dX}
\end{equation}

An important feature of the sigma model that is critical for Pohlmeyer reduction is the fact that the $T_{zz}$ component of the stress-energy tensor is a function solely of $z$, while the $T_{\bar z \bar z}$ component is a function solely of $\bar z$. Indeed, using expression \eqref{eq:Tzz_X} for $T_{zz}$, the equations of motion \eqref{eq:eom_X_1} and the orthogonality between $X^i$ and $\partial X^i$, we find
\begin{equation}
{\bar \partial }{T_{zz}} = \frac{2}{g^2} \left( {\partial }{X^i} \right) \left( {\partial }{\bar \partial }{X^i} \right) =  - \frac{2}{g^2} \lambda \left( {\partial }{X^i} \right) {X^i} = 0 .
\label{eq:Tzz_function_of_z}
\end{equation}
Similarly,
\begin{equation}
{\partial }{T_{\bar z \bar z}} = 0 .
\label{eq:Tzbzb_function_of_zb}
\end{equation}
The stress-energy tensor for meron solutions \eqref{eq:meron_solution_Tzz} and \eqref{eq:meron_solution_Tzbzb}, for elliptic solutions \eqref{eq:elliptic_solution_Tzz} and \eqref{eq:elliptic_solution_Tzbzb}, as well as the fact that the stress-energy tensor for instanton solutions vanishes, are all in accordance with this property.

The above fact implies that there is an appropriate redefinition of the complex coordinate $z$, $z = z \left( w \right)$ and $\bar z = \bar z \left( \bar w \right)$ that allows setting the stress-energy tensor elements to a real and positive constant\footnote{Notice that a conformal transformation that sets the stress-energy tensor components to a constant does not necessarily sets them to a real and positive constant, like in the Lorentzian $O(3)$ NLSM \cite{Pohlmeyer:1975nb}. However, an additional shift of the phase of the complex coordinate always allows such a selection.}
\begin{equation}
g^2 {T_{ww}} = g^2 {T_{\bar w \bar w}} = {\mu ^2}.
\label{eq:Virasoro_constant}
\end{equation}
Notice that \eqref{eq:Virasoro_constant} implies that the fields do not acquire a given value as $\left| w \right| \to \infty$ and thus, such boundary conditions should be abandoned when the Pohlmeyer reduced theory is considered.

In the traditional Pohlmeyer reduction of the Lorentzian $O(3)$ sigma model, the reduced field is defined as the angle between the vectors $\partial_+ X^i$ and $\partial_- X^i$, where $x_\pm$ are light cone coordinates $x_\pm = x_0 \pm x_1$. In this case, it is a natural definition for the reduced degrees of freedom to set ${\partial _+}{X^i}{\partial _-}{X^i} = {\mu ^2}\cos \varphi$, since ${\partial _\pm}{X^i}$ are real vectors with norm equal to $\mu$, as a consequence of the stress-energy tensor being equal to $T_{++} = T_{--} = \mu^2$, similarly to our case. However, in the Euclidean scenario under study, the inner product of $\partial X^i$ and $\bar \partial X^i$ is always positive and larger than the absolute value of $\mu ^2$. If we define the real and imaginary parts of $w$ as $x$ and $y$ and remembering that we have performed an appropriate coordinate redefinition such that $\mu^2$ is real, it is true that
\begin{align}
\mu^2 = \partial {X^i}\partial {X^i} = \bar \partial {X^i}\bar \partial {X^i} &= \frac{1}{4}\left[ {\left( {\frac{{\partial {X^i}}}{{\partial x}}} \right)\left( {\frac{{\partial {X^i}}}{{\partial x}}} \right) - \left( {\frac{{\partial {X^i}}}{{\partial y}}} \right)\left( {\frac{{\partial {X^i}}}{{\partial y}}} \right)} \right] , \\
\partial {X^i}\bar \partial {X^i} &= \frac{1}{4}\left[ {\left( {\frac{{\partial {X^i}}}{{\partial x}}} \right)\left( {\frac{{\partial {X^i}}}{{\partial x}}} \right) + \left( {\frac{{\partial {X^i}}}{{\partial y}}} \right)\left( {\frac{{\partial {X^i}}}{{\partial y}}} \right)} \right] .
\end{align}
Thus, ${\partial }{X^m}{\bar \partial }{X^m} \geq \mu^2$ and a natural definition for the reduced degrees of freedom in the case of the Euclidean sigma model is
\begin{equation}
{\partial }{X^m}{\bar \partial }{X^m} := {\mu ^2}\cosh \varphi .
\label{eq:phi_definition}
\end{equation}
Notice that since ${\partial }{X^m}{\bar \partial }{X^m}$ is both positive and larger than the absolute value of $T_{zz}$, equation \eqref{eq:phi_definition} is a good definition for the reduced field $\varphi$ only if $\mu^2$ has been selected to be positive. Thus, one has to be careful when defining the complex variable $w$, so that $T_{ww}$ is not only constant and real, but also positive.

\subsection{The Basis in Enhanced Space and the Reduced Theory}
\label{subsec:Pohlmeyer_basis}

We would like to specify the dynamics of the field $\varphi$. In order to do so, we need to calculate the derivative ${\partial }{\bar \partial }\cosh \varphi $. Using the definition \eqref{eq:phi_definition} of the field $\varphi$, we get
\begin{multline}
{\mu ^2}{\partial }{\bar \partial }\cosh \varphi = {\partial }{\bar \partial }\left( {{\partial }{X^i}{\bar \partial }{X^i}} \right) \\
 = \partial ^2{X^i}{\bar \partial} ^2{X^i} + \partial ^2{\bar \partial }{X^i}{\bar \partial }{X^i} + {\bar \partial }^2{\partial }{X^i}{\partial }{X^i} + {\partial }{\bar \partial }{X^i}{\partial }{\bar \partial }{X^i}.
 \label{eq:four_terms}
\end{multline}
The first term turns out to be the most complicated to calculate. Finding a base in the enhanced, flat, three-dimensional target space of the theory would be convenient, in order to express the second derivatives of $X^i$ in that base and calculate the desired term.

As explained in section \ref{subsec:Pohlmeyer_definitions}, the vector $X^i$ is perpendicular to the vectors ${\partial }{X^i}$ and ${\bar \partial }{X^i}$. Consequently, unless a degeneracy between the real and imaginary parts of ${\partial }{X^i}$ occurs, the three vectors $X^i$, ${\partial }{X^i}$ and ${\bar \partial }{X^i}$ form a basis in the three-dimensional enhanced flat target space of the sigma model and thus, any vector can be written as a linear combination of the above.

Let's take advantage of this fact and write the second derivative of $X^i$ as such a linear combination,
\begin{equation}
\partial  ^2{X^i} = {a}{X^i} + {b}{\partial }{X^i} + {c}{\bar \partial }{X^i} .
\end{equation}
Then, equation \eqref{eq:X_orthogonal_to_dX} and the definition of the field $\varphi$ \eqref{eq:phi_definition} imply that
\begin{align}
\partial ^2{X^i}{X^i} &= {a} , \\
\partial ^2{X^i}{\partial }{X^i} &= {b}{\mu ^2} + {c}{\mu ^2}\cosh \varphi , \\
\partial ^2{X^i}{\bar \partial }{X^i} &= {b}{\mu ^2}\cosh \varphi  + {c}{\mu ^2} .
\end{align}

The vectors $X^i$ and $\partial X^i$ are orthogonal. It is a direct consequence that
\begin{equation}
{X^i}\partial ^2{X^i} = - {\partial }{X^i}{\partial }{X^i} =  - {\mu ^2} .
\end{equation}
The vector ${\partial }{X^i}$ has constant magnitude and thus,
\begin{equation}
\partial ^2{X^i}{\partial }{X^i} = 0 .
\end{equation}
Finally, from the definition of the field variable $\varphi$, we get
\begin{equation}
\partial ^2{X^i}{\bar \partial }{X^i} =  {\mu ^2}\sinh \varphi {\partial }\varphi .
\end{equation}
Thus, we have all necessary information to specify the coefficients $a$, $b$ and $c$. We find
\begin{align}
{a} &=  - {\mu ^2} ,\\
{b} &=  \coth \varphi {\partial }\varphi , \\
{c} &= - \csch \varphi {\partial }\varphi ,
\end{align}
meaning that
\begin{equation}
\partial  ^2{X^i} = - {\mu ^2}{X^i} + \coth \varphi {\partial }\varphi{\partial }{X^i} - \csch \varphi {\partial }\varphi {\bar \partial }{X^i} .
\label{eq:ddX}
\end{equation}

It is now straightforward to calculate the inner product $\partial ^2{X^i}\partial ^2{X^i}$,
\begin{equation}
\begin{split}
\partial ^2{X^i}{\bar \partial }^2{X^i} &= {a}{\bar a} + {\mu ^2}\left( {{b}{\bar c} + {\bar b}{c}} \right) + {\mu ^2} \left( {{b}{\bar b} + {c}{\bar c}} \right) \cosh \varphi\\
 &= {\mu ^4} + {\mu ^2}\cosh \varphi {\partial }\varphi {\bar \partial }\varphi 
\end{split} .
\end{equation}

The second and third terms of \eqref{eq:four_terms} can be easily calculated with the help of the equations of motion \eqref{eq:eom_X_2}, expressed with the help of field $\varphi$,
\begin{equation}
{\partial }{\bar \partial }{X^i} =  - {\mu ^2}\cosh \varphi {X^i}.
\end{equation}
Differentiating the above and taking the appropriate inner product we find
\begin{equation}
\partial ^2{\bar \partial }{X^i}{\partial }{X^i} = {\bar \partial }^2{\partial }{X^i}{\bar \partial }{X^i} = - {\mu ^4}{\cosh ^2}\varphi .
\end{equation}
Finally, the last term of \eqref{eq:four_terms} can be calculated directly from the equations of motion
\begin{equation}
{\partial }{\bar \partial }{X^i}{\partial }{\bar \partial }{X^i} = {\mu ^4}{\cosh ^2}\varphi .
\end{equation}

All four terms required for the calculation of the derivative ${\partial }{\bar \partial }\cosh \varphi $ through equation \eqref{eq:four_terms} have been specified and we get
\begin{equation}
{\mu ^2}{\partial }{\bar \partial }\cosh \varphi = {\mu ^2}\cosh \varphi {\partial }\varphi {\bar \partial }\varphi - {\mu ^4}{\sinh ^2}\varphi,
\end{equation}
implying that
\begin{equation}
{\partial }{\bar \partial }\varphi  =  - {\mu ^2}\sinh \varphi ,
\end{equation}
which is the sinh-Gordon equation, derivable from the Lagrangian density
\begin{equation}
\mathcal{L} = \frac{1}{2} {\partial }\varphi {\bar \partial }\varphi - {\mu ^2}\cosh \varphi .
\label{eq:reduced_Lagrangian}
\end{equation}

\section{The Non-linear Sigma Model Solutions in the Reduced Formulation}
\label{sec:Reduced_Solutions}

In this section, we find the counterparts of the given solutions of the Euclidean $O(3)$ non-linear sigma model presented in section \ref{sec:NLSM_Solutions}, in the language of the Pohlmeyer reduced theory, namely the sinh-Gordon equation. A basic element of Pohlmeyer reduction, as shown in section \ref{sec:Pohlmeyer}, is the application of a conformal transformation that sets the elements of the stress-energy tensor to a positive constant. Consequently, we will start by performing such a coordinate transformation for each class of solutions.

We would like to point out here that in an obvious way, Pohlmeyer reduction cannot be performed if the stress-energy tensor is vanishing. Thus, it is expected that instanton solutions, which are the only finite-action solutions, do not have a counterpart within the solutions of the sinh-Gordon equation. Consequently, we expect that the counterparts of the elliptic solutions must also become problematic at the $E\to 0$ limit.

\subsection{Meron Solutions in the Reduced Theory}
\label{subsec:Solutions_merons}

The stress-energy tensor elements in the case of meron solutions are given by equations \eqref{eq:meron_solution_Tzz} and \eqref{eq:meron_solution_Tzbzb}. We desire performing a conformal transformation $ z = z\left( w \right) $ so that the stress-energy tensor elements are set to a positive constant. After such a conformal transformation, the new stress-energy tensor $ww$ component is equal to
\begin{equation}
{T_{ww}} =  - \frac{1}{4 g^2} {\left( {\frac{{f'\left( z \right)}}{{f\left( z \right)}}z'\left( w \right)} \right)^2} =  - \frac{1}{4 g^2} {\left( {\frac{{d\ln f}}{{dw}}} \right)^2} .
\end{equation}
So if we demand that $g^2 {T_{ww}} = {\mu ^2}$, the appropriate conformal transformation should obey
\begin{equation}
\frac{{d\ln f}}{{dw}} = 2 i \mu  \Rightarrow f = c{e^{2 i \mu w}} .
\label{eq:meron_conformal_transformation}
\end{equation}

In terms of the new coordinate $w$, the meron solution is written as
\begin{equation}
W = {e^{i\arg c}}{e^{i\mu\left( {w - \bar w} \right)}} .
\end{equation}
Thus, the corresponding solution of the Pohlmeyer reduced theory is
\begin{equation}
{\mu ^2}\cosh \varphi  = \partial X^i \bar \partial X^i = 2\frac{{\partial W\bar \partial \bar W + \partial \bar W\bar \partial W}}{{{{\left( {1 + W\bar W} \right)}^2}}} = {\mu ^2}
\end{equation}
or
\begin{equation}
\varphi  = 0 .
\end{equation}
This implies that all meron solutions are mapped through Pohlmeyer reduction to the ground state of the sinh-Gordon equation.

\subsection{The Elliptic Solutions in the Reduced Theory}
\label{subsec:Solutions_elliptic}

We have seen in Section \ref{subsec:Elliptic}, that the Euclidean $O(3)$ NLSM accepts a family of solutions that correspond to solutions of the one-dimensional pendulum with potential $ V = \cos \psi  - 1$, characterized by the value of the energy integral $E$. The stress-energy tensor for these solutions is given by equations \eqref{eq:elliptic_solution_Tzz} and \eqref{eq:elliptic_solution_Tzbzb}.

\subsubsection{Quasi-Periodic Elliptic Solutions}
As for merons, the reduction of the theory requires the performance of a conformal transformation in order to set the stress-energy tensor equal to a positive constant. For quasi-periodic elliptic solutions, for whom $E>0$, such an appropriate conformal transformation is
\begin{equation}
z = {e^{\frac{w}{a}}} .
\label{eq:elliptic_quasi-periodic_conformal_transformation}
\end{equation}
Then, the stress-energy tensor takes the form
\begin{equation}
{T_{ww}} = \frac{E}{{{8 a^2}}} \equiv {\mu ^2} .
\label{eq:elliptic_quasi-periodic_mu}
\end{equation}

Defining the real and imaginary parts of the complex variable $w$ to be $x$ and $y$ respectively,
\begin{equation}
w = x + iy,
\end{equation}
the elliptic solutions \eqref{eq:elliptic_solution_quasi-periodic} are written as
\begin{equation}
W = {e^{i\frac{y}{a}}}\sqrt {\frac{{1 + \sn\left( {{{\left( {\frac{{E + 2}}{2}} \right)}^{\frac{1}{2}}}\left( {\frac{x}{a} - {t_0}} \right);{{\left( {\frac{2}{{E + 2}}} \right)}^{\frac{1}{2}}}} \right)}}{{1 - \sn\left( {{{\left( {\frac{{E + 2}}{2}} \right)}^{\frac{1}{2}}}\left( {\frac{x}{a} - {t_0}} \right);{{\left( {\frac{2}{{E + 2}}} \right)}^{\frac{1}{2}}}} \right)}}} .
\label{eq:elliptic_quasi-periodic_W_w}
\end{equation}

It turns out that these solutions have a simpler expression in terms of the angular field coordinates $\Theta$ and $\Phi$. Specifically,
\begin{align}
\Theta  &= \am\left( {{{\left( {\frac{{E + 2}}{2}} \right)}^{\frac{1}{2}}}\left( {\frac{x}{a} - {t_0}} \right);{{\left( {\frac{2}{{E + 2}}} \right)}^{\frac{1}{2}}}} \right) - \frac{\pi }{2},\label{eq:elliptic_quasi-periodic_theta}\\
\Phi  &= \frac{y}{a} .\label{eq:elliptic_quasi-periodic_phi}
\end{align}
Since $\Theta$ depends only on the real part of $w$ and $\Phi$ depends only on the imaginary part of $w$, we acquire the following simple formulas using equations \eqref{eq:Tzz_angular} and \eqref{eq:action_angular}
\begin{align}
g^2 T_{ww} = \mu^2 &= \frac{1}{4}\left( {{{\left( {{\partial _x}\Theta } \right)}^2} - {{\sin }^2}\Theta {{\left( {{\partial _y}\Phi } \right)}^2}} \right) ,\\
\mu^2 \cosh \varphi &= \frac{1}{4}\left( {{{\left( {{\partial _x}\Theta } \right)}^2} + {{\sin }^2}\Theta {{\left( {{\partial _y}\Phi } \right)}^2}} \right) .
\end{align}
Thus, the counterpart of the quasi-periodic elliptic solution \eqref{eq:elliptic_solution_quasi-periodic} in the Pohlmeyer reduced theory is
\begin{equation}
\cosh \varphi  = 1 + \frac{{4}}{E}\cn^2\left( {{{\left( {\frac{{E + 2}}{2}} \right)}^{\frac{1}{2}}}\left( {\frac{x}{a} - {t_0}} \right);{{\left( {\frac{2}{{E + 2}}} \right)}^{\frac{1}{2}}}} \right) .
\label{eq:elliptic_quasi-periodic_reduced_field}
\end{equation}

Since the reduced field solution depends only on the real part of the complex variable $w$, the solutions belong to a special subset of the solutions of the Euclidean sinh-Gordon equation that depend on only one real variable, $\varphi = \varphi \left( x \right)$ and consequently they obey
\begin{equation}
\frac{d^2\varphi}{{dx}^2}  =  - 4{\mu ^2}\sinh \varphi .
\end{equation}
These solutions have an one-dimensional mechanical analog of a ``hyperbolic pendulum'' with potential
\begin{equation}
V = 4{\mu ^2}\cosh \varphi ,
\label{eq:reduced_1d_potential}
\end{equation}
which from energy conservation obey
\begin{equation}
E' = \frac{1}{2}{\left( \frac{d\varphi}{dx} \right)^2} + 4{\mu ^2}\cosh \varphi .
\end{equation}
It is a matter of simple algebra to show that equation \eqref{eq:elliptic_quasi-periodic_reduced_field} implies
\begin{equation}
E' = \frac{{E + 4}}{{2{a^2}}} .
\label{eq:elliptic_quasi-periodic_reduced_energy}
\end{equation}

\subsubsection{Periodic Elliptic Solutions}

For periodic elliptic solutions, for whom $E<0$, an appropriate conformal transformation to set the stress-energy tensor to a constant is
\begin{equation}
z = {e^{\frac{w}{i a}}} .
\label{eq:elliptic_periodic_conformal_transformation}
\end{equation}
Then, the stress-energy tensor takes the form
\begin{equation}
{T_{ww}} = - \frac{E}{{{8 a^2}}} \equiv {\mu ^2} .
\label{eq:elliptic_periodic_mu}
\end{equation}

Defining the real and imaginary parts of the complex variable $w$ to be $x$ and $y$ respectively,
\begin{equation}
w = x + iy,
\end{equation}
the elliptic solutions \eqref{eq:elliptic_solution_periodic} are written as
\begin{equation}
W = {e^{ - i\frac{x}{a}}}\sqrt {\frac{{1 + {{\left( {\frac{{E + 2}}{2}} \right)}^{\frac{1}{2}}}\sn\left( {\left( {\frac{y}{a} - {t_0}} \right);{{\left( {\frac{{E + 2}}{2}} \right)}^{\frac{1}{2}}}} \right)}}{{1 - {{\left( {\frac{{E + 2}}{2}} \right)}^{\frac{1}{2}}}\sn\left( {\left( {\frac{y}{a} - {t_0}} \right);{{\left( {\frac{{E + 2}}{2}} \right)}^{\frac{1}{2}}}} \right)}}} .
\label{eq:elliptic_periodic_W_w}
\end{equation}

This implies that the solution in terms of the angular field coordinates is written as
\begin{align}
\Theta  &= \arccos \left[ {{{\left( {\frac{{E + 2}}{2}} \right)}^{\frac{1}{2}}}\sn\left( {\left( {\frac{y}{a} - {t_0}} \right);{{\left( {\frac{{E + 2}}{2}} \right)}^{\frac{1}{2}}}} \right)} \right] ,\label{eq:elliptic_periodic_theta}\\
\Phi  &= - \frac{x}{a} \label{eq:elliptic_periodic_phi}
\end{align}
and similarly to the case of quasi-periodic solutions we can acquire the following simple expressions
\begin{align}
g^2 T_{ww} = \mu^2 &= \frac{1}{4}\left( {{{- \left( {{\partial _y}\Theta } \right)}^2} + {{\sin }^2}\Theta {{\left( {{\partial _x}\Phi } \right)}^2}} \right) ,\\
\mu^2 \cosh \varphi &= \frac{1}{4}\left( {{{\left( {{\partial _y}\Theta } \right)}^2} + {{\sin }^2}\Theta {{\left( {{\partial _x}\Phi } \right)}^2}} \right) .
\end{align}
Thus, the counterpart of the periodic elliptic solution \eqref{eq:elliptic_solution_periodic} in the reduced theory is
\begin{equation}
\cosh \varphi  = \left( {1 - 2\frac{{E + 2}}{E}\cn^2\left( {\left( {\frac{y}{a} - {t_0}} \right);{{\left( {\frac{{E + 2}}{2}} \right)}^{\frac{1}{2}}}} \right)} \right) .
\label{eq:elliptic_periodic_reduced_field}
\end{equation}

As in the case of the quasi-periodic elliptic solution, the reduced field solution depends only on the imaginary part of the complex variable $w$, and, thus, the solutions belong to a special subset of the solutions of the Euclidean  sinh-Gordon equation that depend only on one real variable, $\varphi = \varphi \left( y \right)$ and therefore obey
\begin{equation}
\frac{d^2 \varphi}{{dy}^2}  =  - 4{\mu ^2}\sinh \varphi ,
\end{equation}
being the same ordinary differential equation as in the case of quasi-periodic solutions. Consequently, there is also the same one-dimensional mechanical analog of a ``hyperbolic pendulum'' with potential given by equation \eqref{eq:reduced_1d_potential}. Energy conservation in this one-dimensional mechanical analog implies that the counterparts of periodic elliptic solutions in the Pohlmeyer reduced theory obey
\begin{equation}
E' = \frac{1}{2}{\left( \frac{d \varphi}{dy} \right)^2} + 4{\mu ^2}\cosh \varphi .
\end{equation}
It is a matter of simple algebra to show that equation \eqref{eq:elliptic_periodic_reduced_field} implies that
\begin{equation}
E' = \frac{{E + 4}}{{2{a^2}}} ,
\label{eq:elliptic_periodic_reduced_energy}
\end{equation}
which is the exact same formula as in the case of quasi-periodic solutions.

\subsubsection{Comparison of the Initial and Reduced Formulations}

In the process of deriving the form of the counterparts of the NLSM solutions in the Pohlmeyer reduced theory, we performed a conformal transformation that sets the stress-energy tensor to a real and positive constant. Although equation \eqref{eq:Tzz_function_of_z} ensures that there is always such a transformation, it is not the case that this is the same transformation for all NLSM solutions. This becomes evident comparing the appropriate conformal transformations in the case of meron and elliptic solutions, given by equations \eqref{eq:meron_conformal_transformation}, \eqref{eq:elliptic_quasi-periodic_conformal_transformation} and \eqref{eq:elliptic_periodic_conformal_transformation}. Even in the case of the family of elliptic solutions, the same conformal transformation manages to set the stress-energy tensor components to a real and positive constant for the whole family, but a different one for each member of the family. Since this constant enters into the reduced problem as the mass scale of the sinh-Gordon equation, each solution of the elliptic family has been mapped to a solution of a different version of the sinh-Gordon equation.

\begin{figure}[h]
\begin{center}
\includegraphics[width=0.4\textwidth]{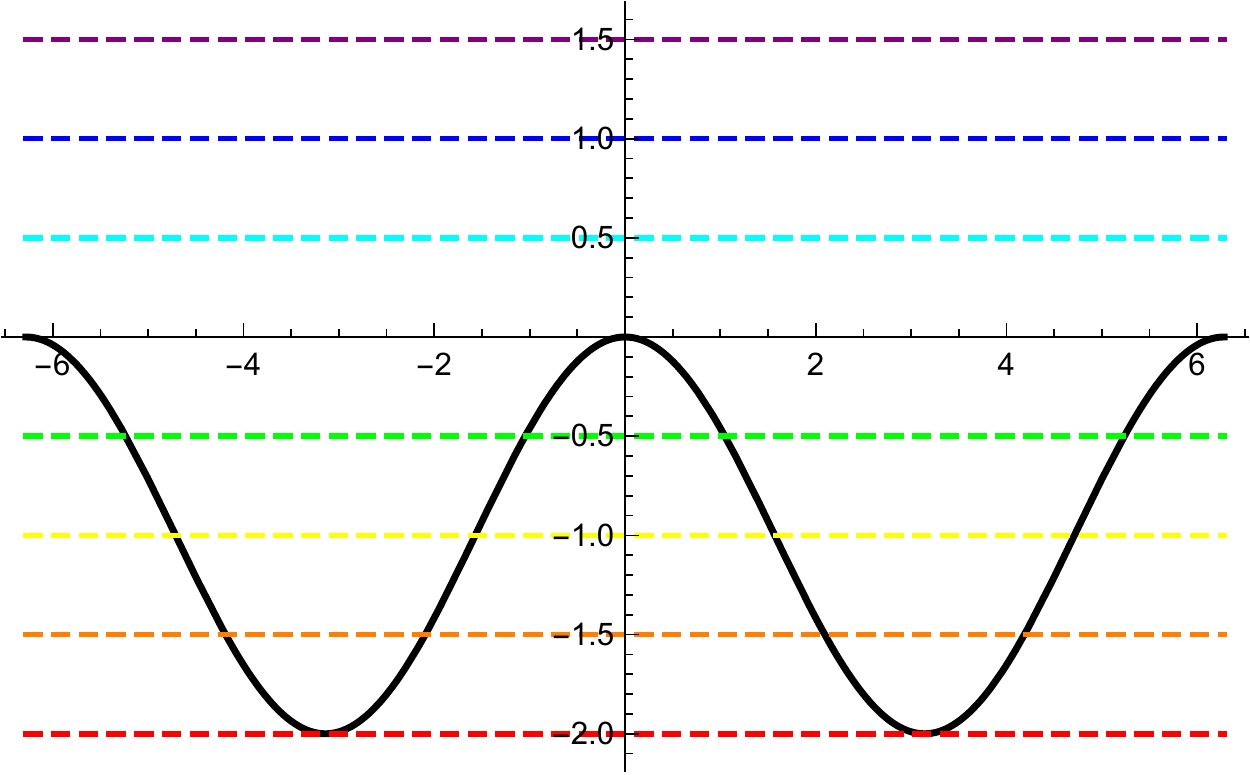}\includegraphics[width=0.4\textwidth]{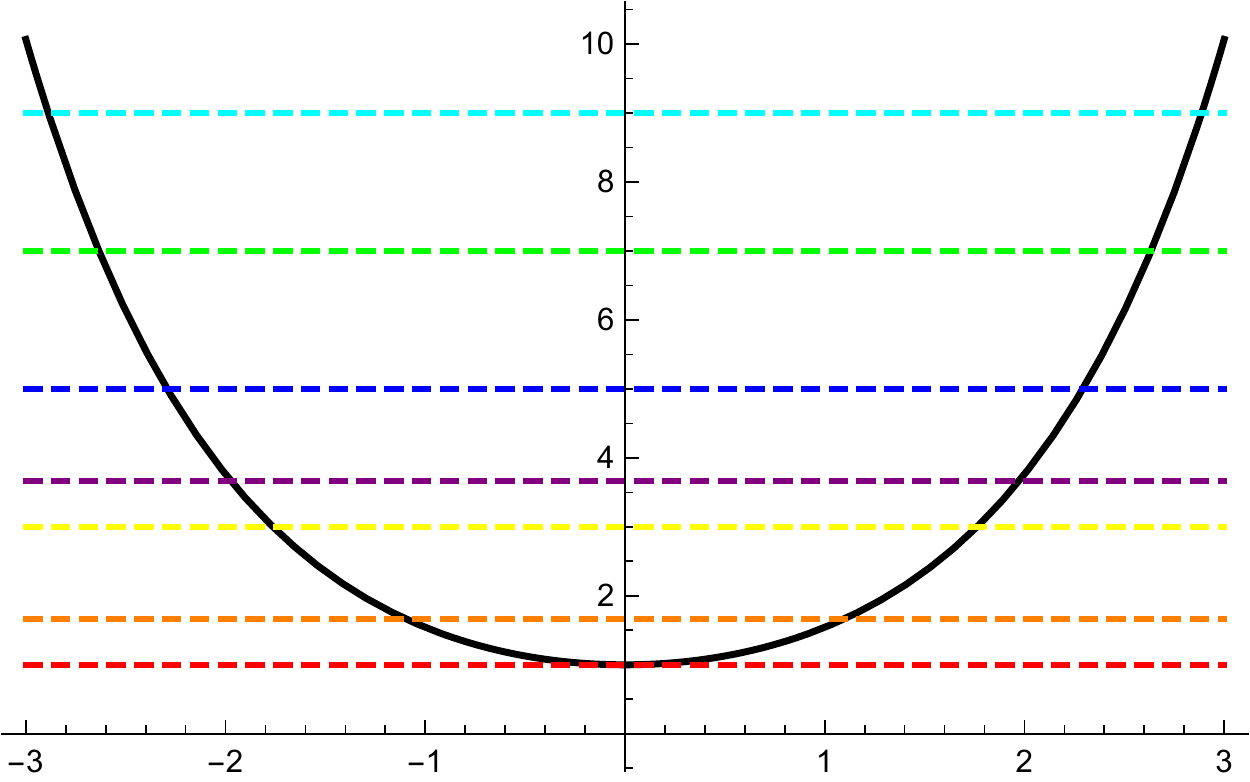}
\end{center}
\caption{The energy levels in the original and reduced problems}
\label{fig:elliptic_comparison}
\end{figure}

In order to better understand the correspondence between the elliptic solutions of the NLSM and the solutions of the sinh-Gordon equation, we would like to make a more specific conformal transformation for each solution, so that the mass scale of the sinh-Gordon equation is always the same. This can be achieved by taking advantage of the freedom to select the parameter $a$ appearing in the conformal transformations \eqref{eq:elliptic_periodic_conformal_transformation} and \eqref{eq:elliptic_quasi-periodic_conformal_transformation}. Choosing
\begin{equation}
{a^2} = \frac{{\left| E \right|}}{2} ,
\label{eq:parameter_a}
\end{equation}
the potential of the one-dimensional effective mechanical problem equals
\begin{equation}
V = \cosh \varphi 
\end{equation}
in all cases. Then, the corresponding energy constant in the reduced one-dimensional mechanical problem (the ``hyperbolic'' pendulum problem) is related with the energy constant in the initial one-dimensional mechanical problem (the pendulum problem) as
\begin{equation}
E' = \frac{{E + 4}}{{\left| E \right|}} .
\label{eq:elliptic_correspondence_energies}
\end{equation}
Notice that $E'$ diverges as $E\to 0$. This is expected since for $E \to 0$ the elliptic solutions tend to the instanton limit, which cannot have a Pohlmeyer counterpart. In figure \ref{fig:elliptic_comparison}, we can see the correspondence of the energy levels between the initial and reduced problems.

\begin{figure}[h]
\begin{center}
\includegraphics[width=0.5\textwidth]{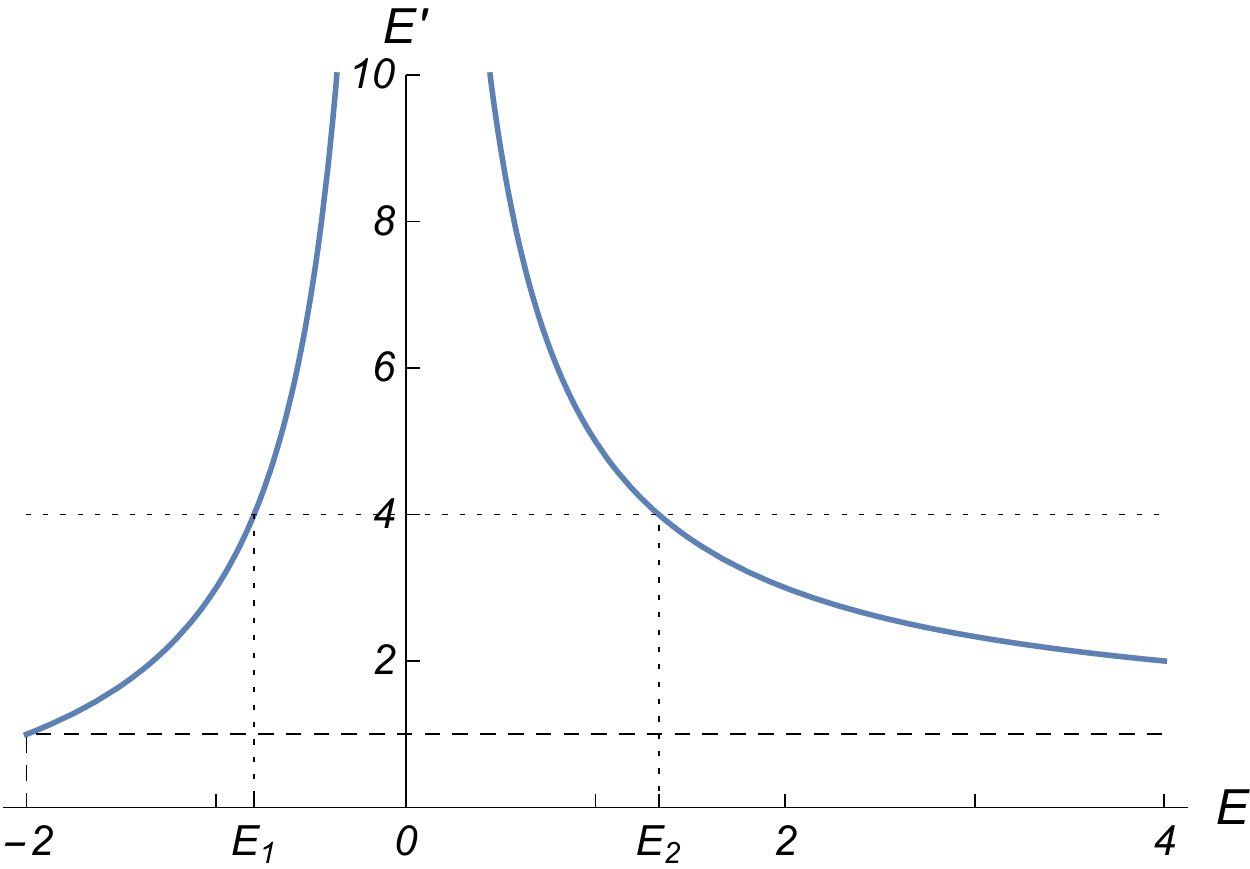}
\end{center}
\caption{The energy constant of the ``hyperbolic pendulum'' solution of the reduced theory as a function of the energy constant of the pendulum solution of the initial formulation of the theory}
\label{fig:energies_relation}
\end{figure}

Interestingly, for every energy level corresponding to a periodic solution of the initial pendulum problem, there is another energy level corresponding to a quasi-periodic solution and vice versa, such that the two solutions correspond to the same energy level in the reduced ``hyperbolic pendulum'' problem. This is evident in figure \ref{fig:energies_relation}, which depicts the relation between the energy levels in the original and reduced problems.
Since the reduced problem solution is characterized only by its energy level (up to a shift in the ``time'' variable), the two solutions of the reduced problem are actually the same solution. One can easily check that equations \eqref{eq:elliptic_quasi-periodic_reduced_energy} and \eqref{eq:elliptic_periodic_reduced_energy}, together with the appropriate setting of the parameter $a$ \eqref{eq:parameter_a}, imply that two energy levels ${E_1} < 0$ and ${E_2} > 0$ in the original problem corresponding to the same energy level in the reduced theory $E'_1 = E'_2$ obey
\begin{equation}
{E_1}{E_2} + 2\left( {{E_1} + {E_2}} \right) = 0 .
\label{eq:elliptic_duality}
\end{equation}
Then, the corresponding solutions in the reduced problem are identical
\begin{equation}
{\varphi ^{\left( p \right)}}\left( {x,y;{E_1}} \right) = {\varphi ^{\left( {qp} \right)}}\left( {y,x;{E_2}} \right) .
\end{equation}
Had we not enforced the modulus of the periodic solutions to lie in the interval $\left[0,1\right]$, as we did to result to formula \eqref{eq:elliptic_solution_periodic} for periodic solutions, but rather we had insisted in describing all solutions with formula \eqref{eq:elliptic_solution_quasi-periodic}, we would have discovered that the moduli of the two solutions with the same counterpart would obey
\begin{equation}
k_1 k_2 = 1
\end{equation}
and thus, the two solutions would be connected with the modular transformation $k \to 1/k$.

On the other hand, the two solutions in the initial formulation of the theory are not identical. The periodic solution corresponds to a surface that covers part of the whole sphere as shown in figure \ref{fig:elliptic_solutions_surfaces}, while the quasi-periodic solution covers the whole sphere, but it is singular at the poles, as the derivatives of the fields do not vanish there. Actually, the discontinuity of the surface corresponding to the periodic solution is similar to the discontinuity to the longitudinal tangential vector of the surface corresponding to the quasi-periodic solution; one can show that
\begin{equation}
{\partial _x}{\Theta ^{\left( {qp} \right)}}\left( {x;{E_2}} \right) \sim \sin {\Theta ^{\left( p \right)}}\left( {x;{E_1}} \right) .
\end{equation}
One may suggest that the two distinct NLSM solutions correspond to the same solution in the reduced theory, as Pohlmeyer reduction deals with the position vector and tangential vector in a symmetric way, as two elements of the basis defined in the enhanced flat three-dimensional target space.

\section{Discussion}
\label{sec:Discussion}

We found the corresponding counterparts of meron and elliptic solutions of Euclidean $O(3)$ non-linear sigma model in the Pohlmeyer reduced version of the same theory, namely the sinh-Gordon equation. Summarizing, we may conclude the following:

Instanton solutions, which are the only finite action solutions of the O$(3)$ model, do not have a counterpart, as they are characterized by vanishing stress-energy tensor. Therefore, not every solution of the non-linear sigma model has a counterpart in the Pohlmeyer reduced theory. Especially solutions with a well defined topological number cannot be mapped to the Pohlmeyer reduced theory, as they have to be characterized by a unique limit at complex infinity, which is inconsistent with the form of the imposed Virasoro constraint in Pohlmeyer reduction.

All meron solutions correspond to the ground state of the sinh-Gordon equation. It seems that this is a more general property that singular configurations such as a meron have. In a similar manner, strings propagating into $R^t \times S^2$ are described by a non-linear sigma model that is reducible to the sine-Gordon equation. Classical string solutions that are singular in a similar sense (the world-sheet tends to an one-dimensional manifold, such as a point particle solution, or the ``hoop'' string solution) have the property of being mapped to either the stable or the unstable vacuum of the sine-Gordon equation \cite{Spradlin:2006wk}.

Elliptic solutions, which can be viewed as solutions of an one-dimensional mechanical pendulum problem in the initial formulation of the theory, have also an one-dimensional mechanical analogue with a hyperbolic cosine potential in the Pohlmeyer reduced description. For every periodic solution in the pendulum problem, there is a quasi-periodic solution and vice versa that has the same counterpart in the hyperbolic cosine problem. The above pair of solutions correspond to physically distinct solutions in the initial formulation of the sigma model.

Furthermore, different, physically inequivalent solutions of the sigma model may correspond to the same solution in the Pohlmeyer reduced theory. This may be attributed to the fact that Pohlmeyer reduction deals with the fields and their derivatives in a symmetric way, as elements of a vector basis in an enhanced higher dimensional space, in our case $\mathbb{R}^3$.

It would be interesting to generalize these results studying the correspondence between solutions of non-linear sigma models with different target space geometries or in higher dimensional target spaces.

\acknowledgments
This work is dedicated to the memory of Prof. Ioannis Bakas. His unexpected loss deprived the scientific community of a member with an elegant way of thinking, infinite organization skills and integrity. This paper is the outcome of uncountable discussions on the connection between geometry and physics and it was built alongside our previous joint work.

\appendix

\section{Useful Formulas for Jacobi's Elliptic Functions}
\label{sec:Elliptic_formulas}

The incomplete elliptic integral of the first kind is defined as
\begin{equation}
F\left( {x;k} \right): = \int_0^x {d\theta {{\left( {1 - {k^2}{{\sin }^2}\theta } \right)}^{ - \frac{1}{2}}}} .
\label{eq:Elliptic_integral}
\end{equation}
The amplitude of Jacobi elliptic functions $\am\left( {x;k} \right)$ is defined as the inverse function of the incomplete elliptic integral of the first kind. Namely if
\begin{equation}
x = F\left( {y;k} \right) ,
\end{equation}
then
\begin{equation}
y: = \am\left( {x;k} \right) .
\label{eq:Jacobi_amplitude}
\end{equation}

The $\sn\left( {x;k} \right)$ and $\cn\left( {x;k} \right)$ Jacobi elliptic functions are defined as the sine and cosine of Jacobi amplitude respectively
\begin{align}
\sn\left( {x;k} \right): &= \sin \left( {\am\left( {x;k} \right)} \right) ,\label{eq:Jacobi_sn}\\
\cn\left( {x;k} \right): &= \cos \left( {\am\left( {x;k} \right)} \right) ,\label{eq:Jacobi_cn}
\end{align}
while the third basic Jacobi elliptic function $\dn\left( {x;k} \right)$ is defined as
\begin{equation}
\dn\left( {x;k} \right): = {\left( {1 - {k^2}{\kern 1pt} \sn^2\left( {x;k} \right)} \right)^{\frac{1}{2}}} .
\label{eq:Jacobi_dn}
\end{equation}

Following from the definitions, the Jacobi elliptic functions have the following properties
\begin{align}
\frac{d}{{dx}}\am\left( {x;k} \right) &= \dn\left( {x;k} \right) ,\label{eq:Jacobi_am_derivative}\\
\frac{d}{{dx}}\sn\left( {x;k} \right) &= \cn\left( {x;k} \right)\dn\left( {x;k} \right) ,\label{eq:Jacobi_sn_derivative}\\
\frac{d}{{dx}}\cn\left( {x;k} \right) &=  - \sn\left( {x;k} \right)\dn\left( {x;k} \right) ,\label{eq:Jacobi_cn_derivative}\\
\frac{d}{{dx}}\dn\left( {x;k} \right) &=  - {k^2}{\kern 1pt} \sn\left( {x;k} \right)\cn\left( {x;k} \right) .\label{eq:Jacobi_dn_derivative}
\end{align}

Jacobi elliptic functions take the following special values
\begin{align}
\sn\left( {0;k} \right) &= 0 ,\label{eq:zero_argument_sn}\\
\cn\left( {0;k} \right) &= 1 ,\label{eq:zero_argument_cn}\\
\dn\left( {0;k} \right) &= 1\label{eq:zero_argument_dn}
\end{align}
and
\begin{align}
\sn\left( {x;1} \right) &= \tanh x ,\label{eq:unit_modulus_sn}\\
\cn\left( {x;1} \right) &= \sech x ,\label{eq:unit_modulus_cn}\\
\dn\left( {x;1} \right) &= \sech x .\label{eq:unit_modulus_dn}
\end{align}

Moreover, Jacobi functions, as they are elliptic functions, are doubly periodic functions, which obey
\begin{align}
\sn\left( {x + 2mK\left( k \right) + 2nK\left( k' \right) ;k} \right) &= {\left( { - 1} \right)^m}\sn\left( {x;k} \right) ,\label{eq:sn_period}\\
\cn\left( {x + 2mK\left( k \right) + 2nK\left( k' \right) ;k} \right) &= {\left( { - 1} \right)^{m+n}}\cn\left( {x;k} \right) ,\label{eq:cn_period}\\
\dn\left( {x + 2mK\left( k \right) + 2nK\left( k' \right) ;k} \right) &= {\left( { - 1} \right)^n}\dn\left( {x;k} \right) ,\label{eq:dn_period}
\end{align}
where $K\left( k \right)$ is the complete elliptic integral $K\left( k \right) = F\left( \pi / 2 ; k \right)$ of the first kind and $k' = \sqrt{1 - k^2}$.

By their definition, Jacobi elliptic functions are real only for real modulus $k$ taking values in the interval $\left[0,1\right]$. However, the Jacobi elliptic functions with real moduli greater that one can be connected with ones with moduli in the interval $\left[0,1\right]$ with the following relations
\begin{align}
\sn\left( {z;\frac{1}{k}} \right) &= k\sn\left( {\frac{z}{k};k} \right) ,\label{eq:sn_reciprocal}\\
\cn\left( {z;\frac{1}{k}} \right) &= \dn\left( {\frac{z}{k};k} \right) ,\label{eq:cn_reciprocal}\\
\dn\left( {z;\frac{1}{k}} \right) &= \cn\left( {\frac{z}{k};k} \right) .\label{eq:dn_reciprocal}
\end{align}

\pagebreak

\end{document}